\documentclass[useAMS,usenatbib]{mn2e}
\usepackage{psfrag,graphicx}
\usepackage{color}
\usepackage{txfonts}
\usepackage{breqn}
\usepackage[T1]{fontenc}
\usepackage{ae,aecompl}
\usepackage{graphicx}
\title[Black hole formation in the early universe]
   {Magnetic fields during the formation of supermassive black holes}
\author[Latif et al.]
  {M.~A.~Latif,$^1$
  D.~R.~G.~Schleicher,$^1$ 
  W.~Schmidt$^1$
   \newauthor 
   $^1$ Institut f\"ur Astrophysik, Georg-August-Universit\"at, \\
    Friedrich-Hund-Platz 1, D-37077 G\"ottingen, Germany}
\date{}

\pagerange{\pageref{firstpage}--\pageref{lastpage}} \pubyear{2009}

\def\LaTeX{L\kern-.36em\raise.3ex\hbox{a}\kern-.15em
      T\kern-.1667em\lower.7ex\hbox{E}\kern-.125emX}

\begin{document}

\bibliographystyle{mn2e}

\label{firstpage}

\maketitle
\def\na{NewA}%
\def\aj{AJ}%
\def\araa{ARA\&A}%
\def\apj{ApJ}%
\def\apjl{ApJ}%
\def\apjs{ApJS}%
\def\ao{Appl.~Opt.}%
\def\apss{Ap\&SS}%
\def\aap{A\&A}%
\def\aapr{A\&A~Rev.}%
\def\aaps{A\&AS}%
\def\azh{AZh}%
\def\baas{BAAS}%
\def\jrasc{JRASC}%
\def\memras{MmRAS}%
\def\mnras{MNRAS}%
\def\pra{Phys.~Rev.~A}%
\def\prb{Phys.~Rev.~B}%
\def\prc{Phys.~Rev.~C}%
\def\prd{Phys.~Rev.~D}%
\def\pre{Phys.~Rev.~E}%
\def\prl{Phys.~Rev.~Lett.}%
\def\pasp{PASP}%
\def\pasj{PASJ}%
\def\qjras{QJRAS}%
\def\skytel{S\&T}%
\def\solphys{Sol.~Phys.}%

\def\sovast{Soviet~Ast.}%
\def\ssr{Space~Sci.~Rev.}%
\def\zap{ZAp}%
\def\nat{Nature}%
\def\iaucirc{IAU~Circ.}%
\def\aplett{Astrophys.~Lett.}%
\def\apspr{Astrophys.~Space~Phys.~Res.}%
\def\bain{Bull.~Astron.~Inst.~Netherlands}%
\def\fcp{Fund.~Cosmic~Phys.}%
\def\gca{Geochim.~Cosmochim.~Acta}%
\def\grl{Geophys.~Res.~Lett.}%
\def\jcp{J.~Chem.~Phys.}%
\def\jgr{J.~Geophys.~Res.}%
\def\jqsrt{J.~Quant.~Spec.~Radiat.~Transf.}%
\def\memsai{Mem.~Soc.~Astron.~Italiana}%
\def\nphysa{Nucl.~Phys.~A}%
\def\physrep{Phys.~Rep.}%
\def\physscr{Phys.~Scr}%
\def\planss{Planet.~Space~Sci.}%
\def\procspie{Proc.~SPIE}%

%


\begin{abstract}
{ Observations of quasars at $\rm z> 6$ report the existence of a billion solar mass black holes. Comprehending their formation in such a short time scale is a matter of ongoing research. One of the most promising scenarios to assemble supermassive black holes is a monolithic collapse of protogalactic gas clouds in atomic cooling halos with $\rm T_{vir} \geq 10^{4}~K$. In this article, we study the amplification and impact of magnetic fields during the formation of seed black holes in massive primordial halos. We perform high resolution cosmological magnetohydrodynamics simulations for four distinct halos and follow their collapse for a few free-fall times until the simulations reach a peak density of $\rm 7 \times 10^{-10}~g/cm^{3}$. Our findings show that irrespective of the initial seed field, the magnetic field strength reaches a saturated state in the presence of strong accretion shocks. Under such conditions, the growth time becomes very short and amplification occurs rapidly within a small fraction of the free-fall time. We find that the presence of such strong magnetic fields provides additional support against gravity and helps in suppressing fragmentation. Massive clumps of a few hundred solar masses are formed at the end of our simulations and high accretion rates of $\rm 1~M_{\odot}/yr$ are observed. We expect that in the presence of such accretion rates, the clumps will grow to form supermassive stars of $\rm \sim 10^{5}~M_{\odot}$. Overall, the role of the magnetic fields seems supportive for the formation of massive black holes.

}
\end{abstract}


\begin{keywords}
methods: numerical -- cosmology: theory -- early Universe -- galaxies: formation
\end{keywords}

\section{Introduction}

Supermassive black holes of about a billion solar masses have been observed at $z > 6$ \citep{2003AJ....125.1649F,2006AJ....131.1203F,2011Natur.474..616M}. How such massive objects are assembled within a billion years after the Big Bang remains one of the unresolved mysteries in the Universe.

Various pathways have been suggested to explain the formation of supermassive black holes in the early universe \citep{1984ARA&A..22..471R,2008arXiv0803.2862D,2009ApJ...702L...5B,2009MNRAS.396..343R,2010A&ARv..18..279V,2012arXiv1203.6075H,2012ApJ...750...66J,2013ApJ...771..116J,2013arXiv1309.1067S} such as merging and accretion of PopIII remnants \citep{2001ApJ...552..459H,2004ApJ...613...36H,2009ApJ...696.1798T,2012ApJ...756L..19W,2013ApJ...772L...3L}, collapse of a dense stellar cluster \citep{2004Natur.428..724P,2008ApJ...686..801O, 2009ApJ...694..302D} and the monolithic collapse of a massive primordial gas cloud \citep{2002ApJ...569..558O,2003ApJ...596...34B,2006ApJ...652..902S,2006MNRAS.370..289B,2006MNRAS.371.1813L,2008MNRAS.391.1961D,2008arXiv0803.2862D,2010MNRAS.402.1249S,2010MNRAS.tmp.1427J,2010ApJ...712L..69S,2011MNRAS.411.1659L,2013arXiv1304.1369C,2013MNRAS.433.1607L,2013ApJ...774...64W,2013arXiv1309.1097L}. The necessary conditions for the direct collapse model are that the gas must be of a primordial composition and the formation of molecular hydrogen remains inhibited. The latter can be achieved in the presence of a strong Lyman Werner flux produced by the stellar populations in the first galaxies \citep{2001ApJ...546..635O,2007MNRAS.374.1557J,2008MNRAS.391.1961D,2010MNRAS.402.1249S,2011MNRAS.410..919J,2010ApJ...712L..69S,2011MNRAS.418..838W,2011A&A...532A..66L,2012MNRAS.425.2854A,2013MNRAS.430..588L}, also see \cite{2012MNRAS.422.2539I,2013A&A...553L...9V}. The potential sites for the direct collapse are the massive primordial halos of $\rm 10^{7}-10^{8}~M_{\odot}$ where the above mentioned conditions can be fulfilled. Numerical simulations performed to study the collapse of a protogalactic halo in the presence of a strong Lyman Werner flux show that massive objects can be formed  \citep{2003ApJ...596...34B,2008ApJ...682..745W,2009MNRAS.393..858R,2011MNRAS.411.1659L,2013MNRAS.433.1607L}. Furthermore, theoretical models propose that supermassive stars formed as a result of direct collapse are the potential embryos of supermassive black holes  \citep{2008MNRAS.387.1649B,2010MNRAS.402..673B,2011MNRAS.414.2751B,2012ApJ...756...93H,2012MNRAS.421.2713B,2013A&A...558A..59S,2013ApJ...768..195W,2013arXiv1308.4457H}.

Previous numerical simulations  mainly focused on the hydrodynamics of the problem, while the role of magnetic fields during the formation of seed black holes via direct collapse remained largely unexplored. Magnetic fields are expected to influence the formation of black holes by exerting extra magnetic pressure and providing additional means for the transport of angular momentum by magnetic torques. Magnetic pressure may enhance the Jeans mass ($ M_{J,B} \propto B^{3}/ \rho^{2}$) and consequently help in suppressing fragmentation which is a key requirement for the direct collapse model. The role of magnetic torques is expected to become significant in the central accretion disk, implying the presence of strong rotation measures and enhanced accretion rates. In fact, the detection of strong rotation measures in quasars at $z = 5.3$ indicates the relevance of magnetic fields in the early universe \citep{2012arXiv1209.1438H}. It is further known from observations of nearby active galactic nuclei that magnetic fields play a vital role in the transport of angular momentum \citep{1999Natur.397..324B,Beck05}.


The standard model of cosmology does not provide any constraints on the initial magnetic fields strength. They could be generated via electro-weak or quantum chromodynamics phase transitions \citep{1996PhRvD..53..662B,1989ApJ...344L..49Q} or alternatively, during structure formation via mechanisms such as the Biermann battery effect, the Weibel instability \citep{1950ZNatA...5...65B,1959PhRvL...2...83W,2003ApJ...599L..57S} or thermal fluctuations in the plasma \citep{2012PhRvL.109z1101S}. In addition to the gravitational compression under the constraint of flux freezing, astrophysical dynamos can efficiently amplify the magnetic field, particularly the small scale dynamo which operates by converting the turbulent energy into the magnetic energy \citep{1968JETP...26.1031K,2005PhR...417....1B,Schobera,2013NJPh...15b3017S}. Numerous studies confirm that the small scale dynamo gets excited during structure formation provided that turbulent energy is well resolved   \citep{2010A&A...522A.115S,2011ApJ...731...62F,2010ApJ...721L.134S,Schobera,2012ApJ...745..154T,Schoberb,2013NJPh...15a3055B,2013MNRAS.432..668L}. The amplification of magnetic fields by the small scale dynamo was further confirmed by \cite{Federrath11} for higher Mach numbers and by \cite{2012ApJ...760L..28P} for different thermodynamical conditions. A recent study by \cite{2013MNRAS.tmp.2194M} shows that the magnetic field may have a significant impact on the formation of Pop III stars as it strongly influences the fragmentation properties of a gas cloud.  In the context of black hole formation via direct collapse, we have shown in our previous study \citep{2013MNRAS.432..668L} that for a Jeans resolution of 64 cells, the small scale dynamo gets excited and exponentially amplifies the magnetic field. It is thus expected that magnetic fields can influence the formation of seed black holes. A recent study by \cite{2014ApJ...782..108S} further shows that the radiation source can aid the generation of magnetic fields.

In this study, we explore for the first time  the impact of magnetic fields on the fragmentation properties of atomic cooling halos, the potential birthplaces of supermassive black holes. To accomplish this goal, we perform high resolution cosmological magnetohydrodynamical simulations for four distinct halos and employ a fixed resolution of 64 cells per Jeans length during the entire course of the simulations. To investigate the impact of saturated magnetic fields on fragmentation, the initial seeds of higher magnetic field strength are selected based on the results of our previous study \citep{2013MNRAS.432..668L}. We employ a constant background Lyman Werner flux  of strength $\rm 10^3$ in units of $\rm J_{21}$ and follow the collapse for a few free fall times by evolving the simulations beyond the formation of the first peak. This study enables us to assess the role of magnetic fields in the assembling of supermassive black holes via direct collapse.

The organization of this article is the following. In the second section, we describe the numerical methods and simulation setup employed in this work. The main results from this study are presented in the third section. We discuss our conclusion and summary of the main findings in the fourth section.

\section{Computational methods}
The simulations presented here are performed with the publicly available cosmological magnetohydrodynamics code ENZO \citep{2004astro.ph..3044O,2013arXiv1307.2265T}. It is a massively parallel code and very well suited for simulations following the collapse from cosmological scales down to the scales of AU. The equations of magnetohydrodynamics (MHD) are solved employing the  Harten-Lax-van Leer (HLL) Riemann solver with a piece-wise linear construction. The Dedner scheme \citep{2008ApJS..176..467W,2010NewA...15..581W} is imposed for divergence cleaning.

We start our simulations at $ z=$ 100 with cosmological initial conditions which are generated using the inits package. Our computational volume has a comoving size of 1~$\rm Mpc/h$ and periodic boundary conditions are employed both for magneto-hydrodynamics and gravity. We initially run  $\rm 128^3$ cells uniform grid simulations to select the most massive halos forming in our computational domain for various random seeds. Simulations are restarted with two additional nested refinement levels each with a resolution of $\rm 128^3$ cells centered on the most massive halo. To simulate the evolution of dark matter dynamics, 5767168  particles are initialized  which provide a particle mass resolution of $\rm \sim 600~M_{\odot}$. During the course of the simulations, additional 27 dynamic refinement levels are employed which yield an effective resolution of 0.25 AU. Apart from the fixed Jeans resolution of 64 cells, our resolution criteria are based on the gas over-density and the particle mass resolution. The cells exceeding four times the mean baryonic density are marked for refinement. Similarly, grid cells are flagged for refinement if the dark matter density exceeds 0.0625 times  $ \rho_{DM}r^{l (1+ \alpha)}$ where $r=$ 2 is the refinement factor, $l$ is the refinement level and $\alpha =-0.3$ makes the refinement super-Lagrangian. Although gravity by the baryons dominates in the core of the simulated halos, the smoothing of dark matter particles becomes essential in order to avoid a spurious heating of the baryons. We smooth particles at scales of 0.68 pc which corresponds to a refinement level of 14. Our approach is similar to the simulations performed to explore gravitational collapse in previous studies \citep{2008ApJ...682..745W,2012ApJ...745..154T,2013MNRAS.433.1607L,2013ApJ...772L...3L}.

The simulations are evolved adiabatically above densities of $\rm 10^{-11}~g/cm^{3}$ after reaching the maximum refinement level to follow the collapse for several dynamical times. Such an approach makes the structures stable on the smallest scales while collapse proceeds on larger scales. We consider these adiabatic cores as proxies for supermassive protostars, which are expected to form at higher densities where cooling is suppressed by the continuum opacity \citep{2001ApJ...546..635O,2008ApJ...686..801O}. In total, we perform eight simulations for four distinct halos each with a weak and a strong initial seed field. The strength of the initial seed fields and the properties of the halos are listed in table \ref{table1}. The simulations are compared at a peak density of $\rm 7 \times 10^{-10}~g/cm^3$. Similar to our previous studies \citep{2013MNRAS.432..668L,2013MNRAS.430..588L}, we employ a strong Lyman Werner flux of strength $\rm 10^3$ in units of $\rm J_{21}=~erg~cm^{-2}~s^{-1}~Hz^{-1}~sr^{-1}$  for stellar spectra of $\rm 10^5$~K and ignore the effect of self-shielding. To model the thermal evolution of the gas, the rate equations of $\rm H$,~$\rm H^{+}$,~$\rm He$,~$\rm He^{+}$,~$\rm He^{++}$,~$\rm e^{-}$,~$\rm H^{-}$,~$\rm H_{2}$,~$\rm H_{2}^{+}$ are self consistently solved with cosmological simulations. 

\begin{table*}
\begin{center}
\caption{Properties of the simulated halos are listed here.}
\begin{tabular}{cccccc}
\hline
\hline

Model	& Initial Mass			& spin parameter     & Collapse redshift    & Initial Magnetic field strength, & Fragmentation  \\

 & $\rm M_{\odot} $	 & $\lambda$    & z  & [$\rm Gauss $] & (For unsaturated cases) \\ 
\hline                                                          \\
 		
 A	  & $\rm 4.3 \times 10^{6}$	& 0.0309765 	&11.3  & $\rm 3 \times 10^{-20}$, $\rm 3 \times 10^{-11}$ & No\\
 B        & $\rm 1.0 \times 10^{7}$    & 0.0338661      &12.8  & $\rm 3 \times 10^{-20}$, $\rm 3 \times 10^{-11}$ & Yes\\
 C        & $\rm 2.3 \times 10^{7}$     & 0.021782      &15.9  & $\rm 3 \times 10^{-20}$, $\rm 3 \times 10^{-11}$ & Yes\\
 D        & $\rm 1.9 \times 10^{7}$     & 0.0084786     &13.7  & $\rm 3 \times 10^{-20}$, $\rm 3 \times 10^{-11}$ & No\\
 
\hline
\end{tabular}
\label{table1}
\end{center}

\end{table*}

\begin{figure*}
\hspace{-10.0cm}
\centering
\begin{tabular}{c}
\begin{minipage}{6cm}
\includegraphics[scale=0.2]{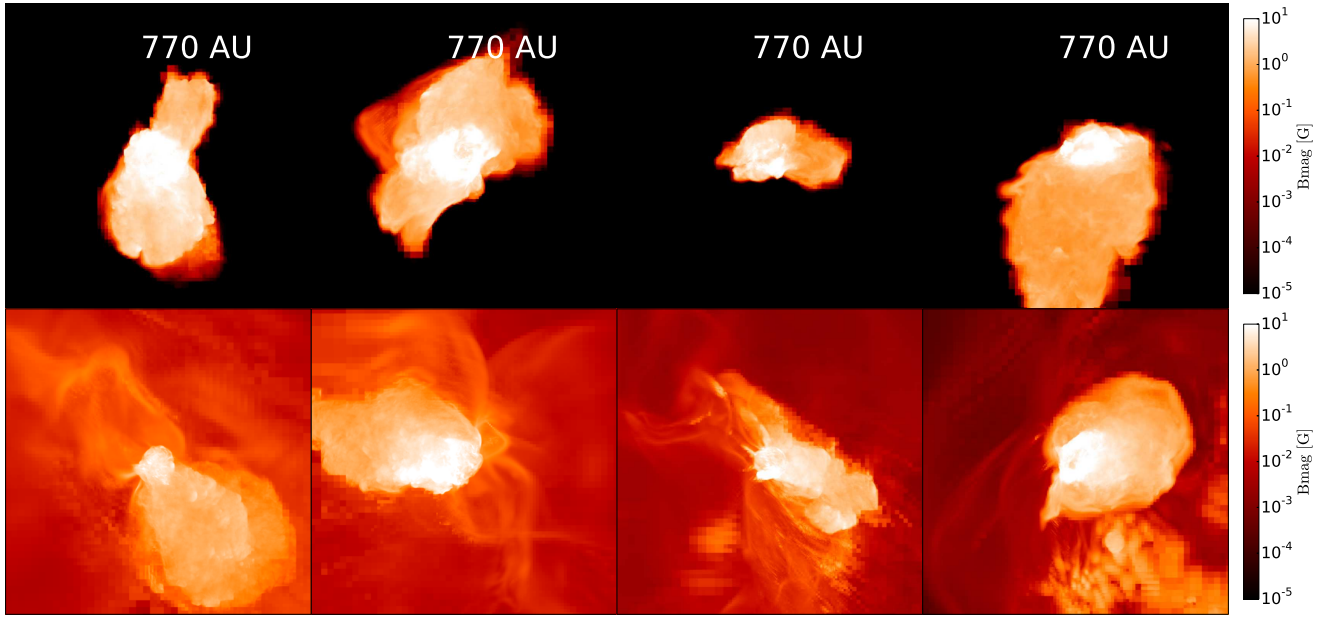}
\end{minipage}
\end{tabular}
\caption{This figure shows the density-weighted magnetic field strength for four halos at the end of our simulations. The top panels show the non-saturated cases while bottom panels depict the saturated cases. Both panels from left to right show the halos A to D as listed in table \ref{table1}.}
\label{figh1}
\end{figure*}

\begin{figure*}
\centering
\begin{tabular}{c c}
\begin{minipage}{4cm}
\hspace{-4cm}
\includegraphics[scale=0.4]{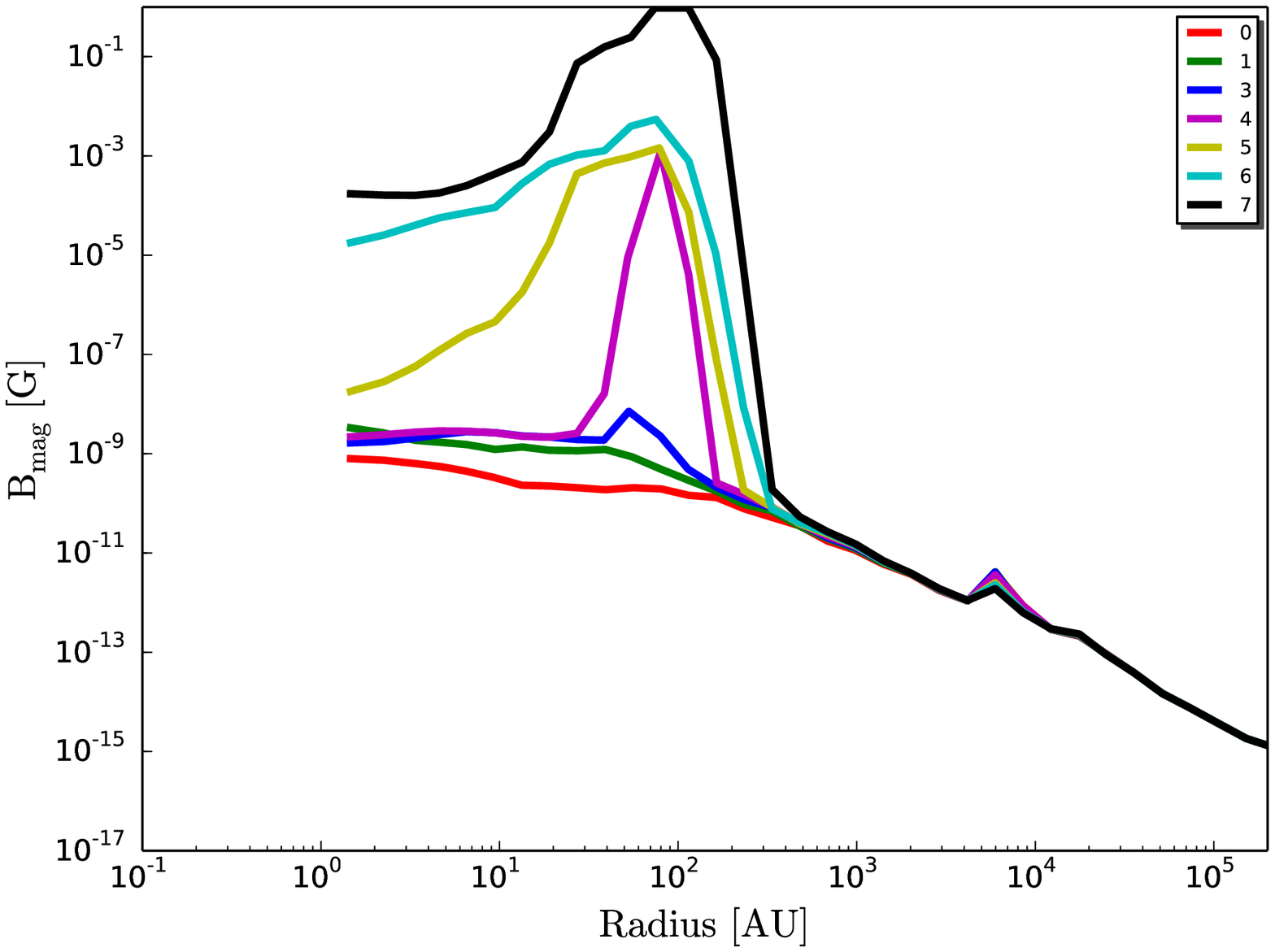}
\end{minipage} &
\begin{minipage}{4cm}
\includegraphics[scale=0.4]{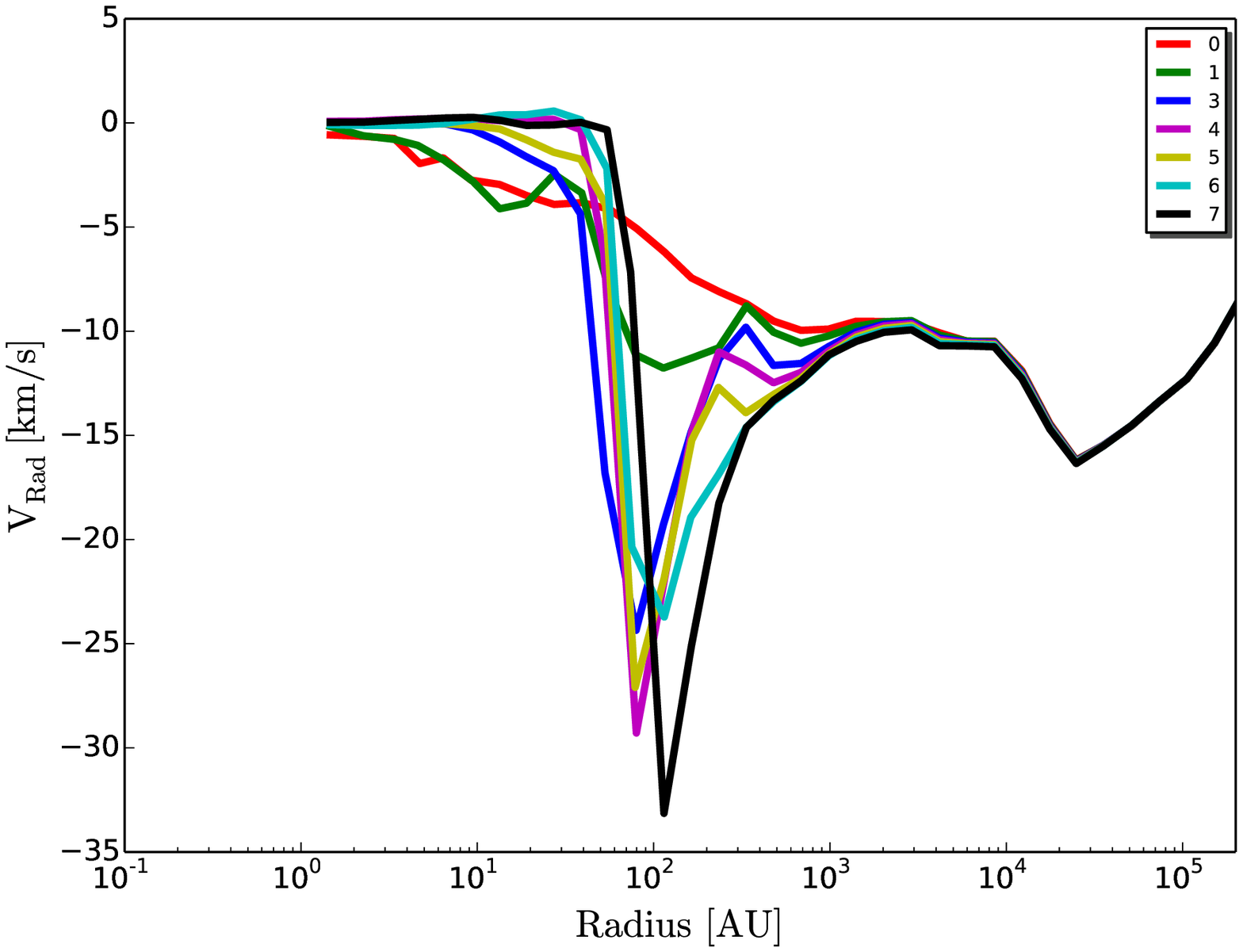}
\end{minipage} 
\end{tabular}
\caption{Time evolution of the magnetic field strength and radial velocity is shown for halo A, the unsaturated case. Each line color represents different time evolution in units of years as mentioned in the legend. }
\label{fig2}
\end{figure*}

\begin{figure*}
\centering
\begin{tabular}{c c}
\begin{minipage}{4cm}
\hspace{-4cm}
\includegraphics[scale=0.4]{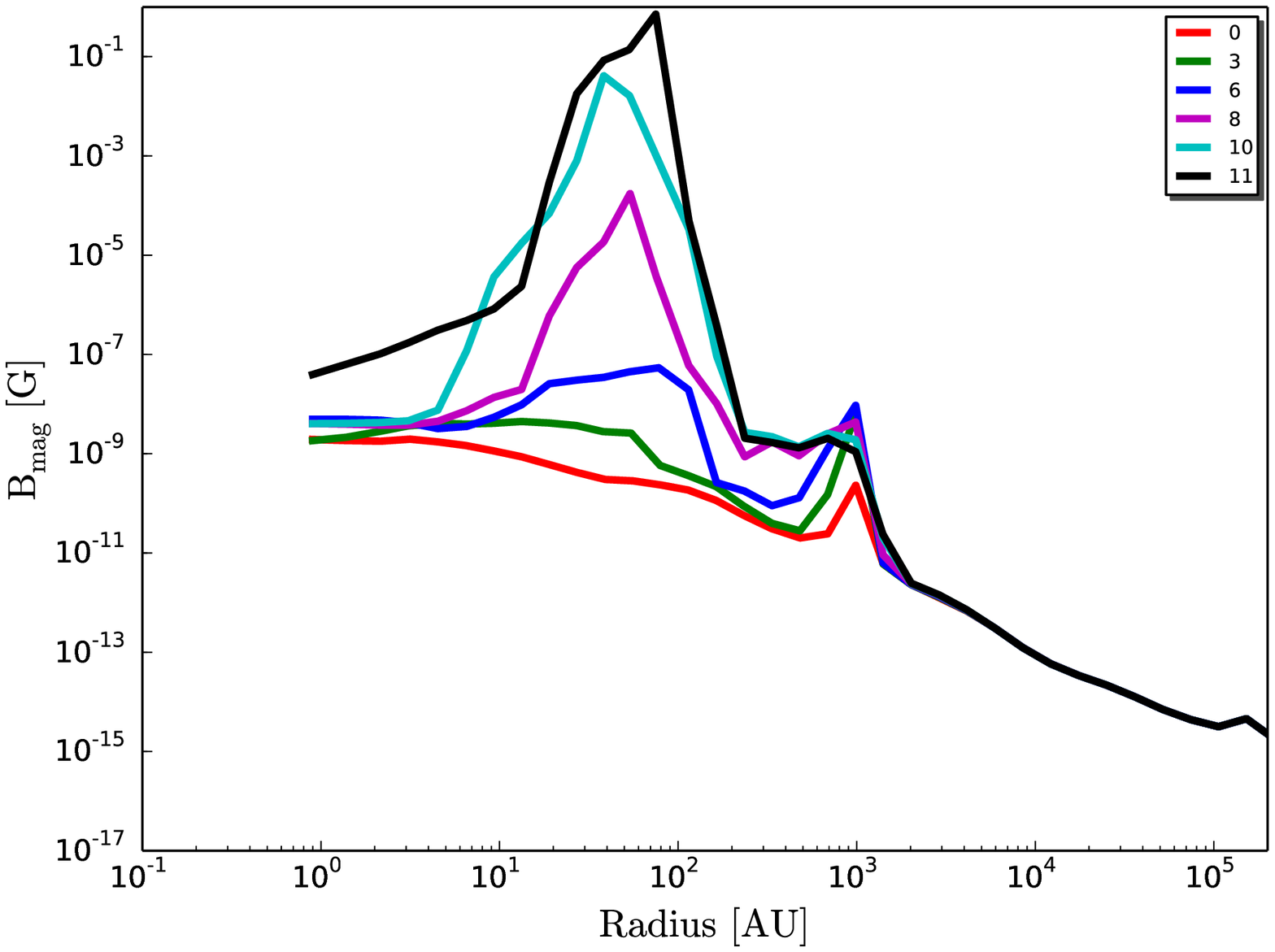}
\end{minipage} &
\begin{minipage}{4cm}
\includegraphics[scale=0.4]{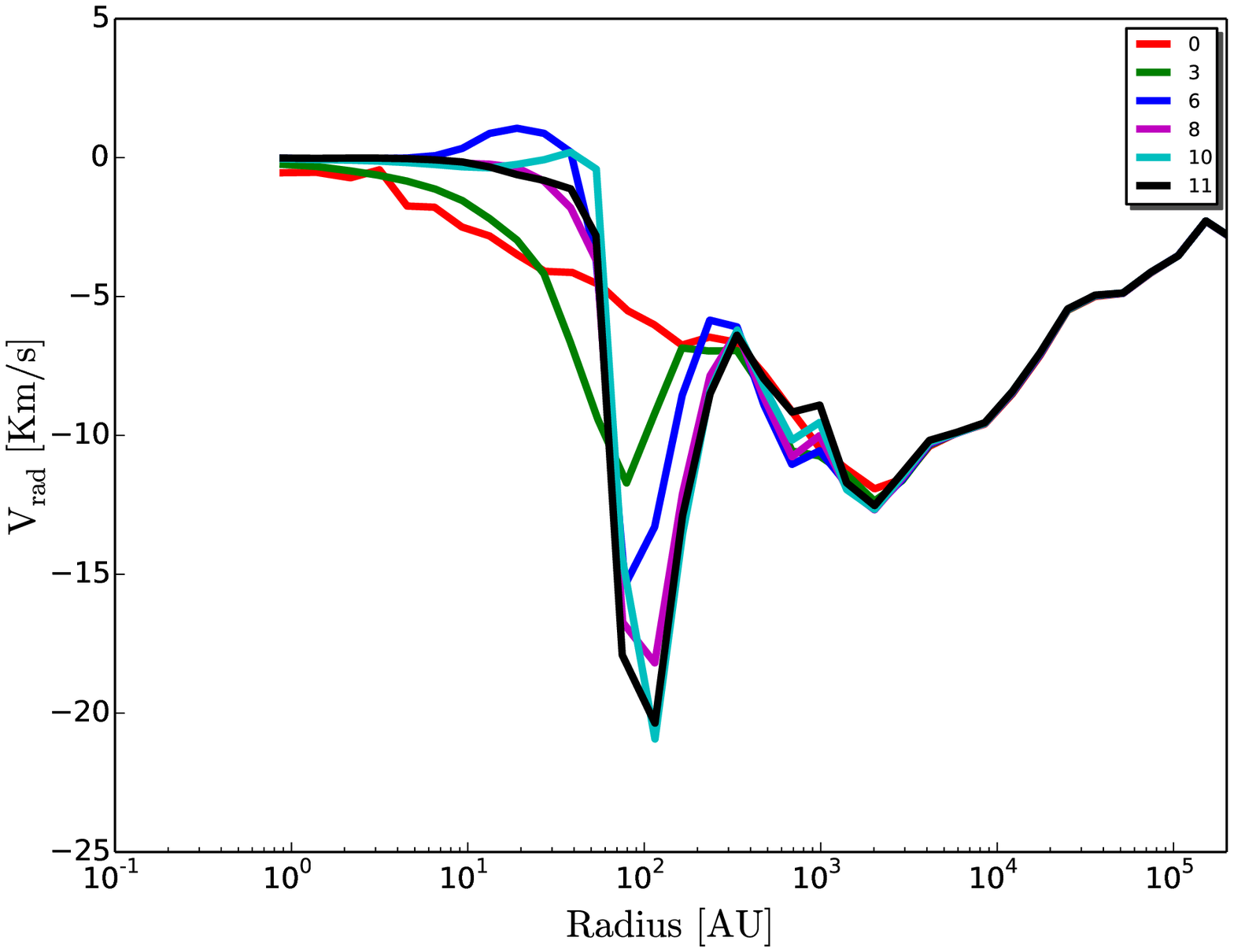}
\end{minipage} 
\end{tabular}
\caption{Same as figure \ref{fig2}. Here we show the time evolution of these quantities for halo C, the unsaturated case.}
\label{fig3}
\end{figure*}

\begin{figure*}
\centering
\begin{tabular}{c c}
\begin{minipage}{4cm}
\hspace{-4cm}
\includegraphics[scale=0.4]{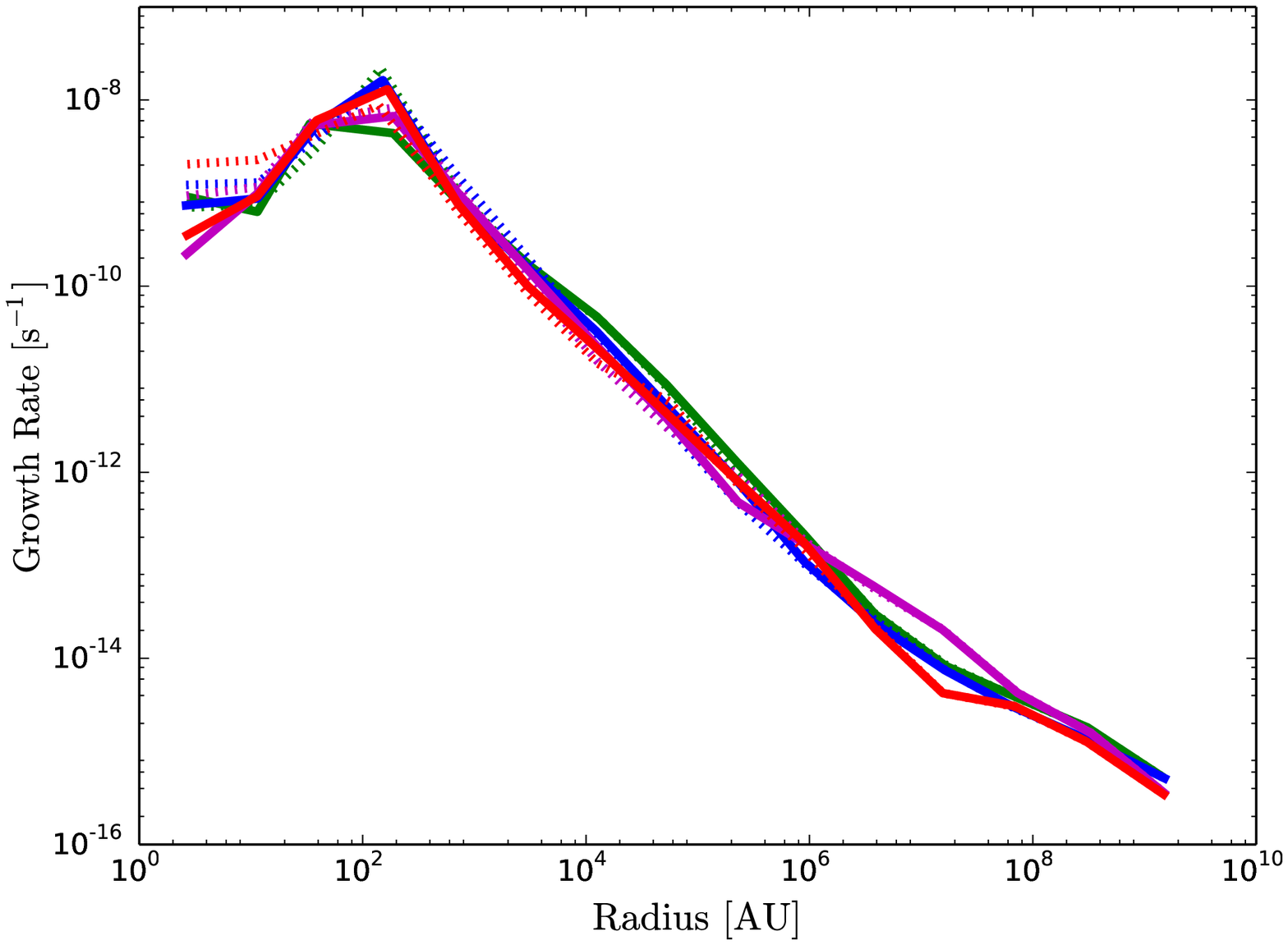}
\end{minipage} &
\begin{minipage}{4cm}
\includegraphics[scale=0.4]{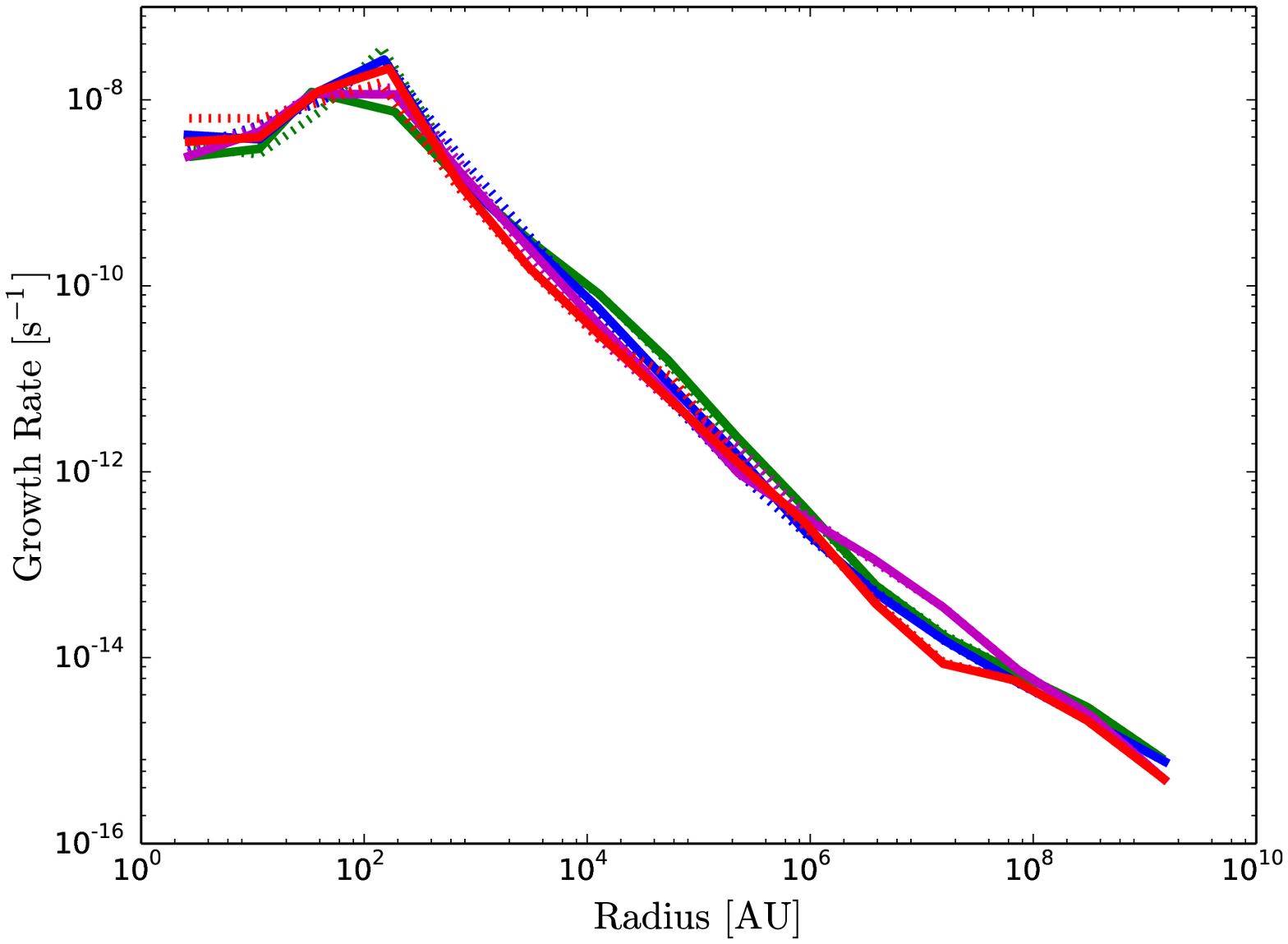}
\end{minipage} 
\end{tabular}
\caption{The absolute values of the growth rates of magnetic field amplification of all halos are plotted against the radius in this figure. The left panel shows the magnetic growth rate due to the compression while right panel shows the  growth rate due to the shear. The dotted lines present unsaturated cases while the solid lines stand for the saturated cases. For the definition of growth rate see the text and references therein. Each color represents a halo. }
\label{fig5}
\end{figure*}

\begin{figure*}
\centering
\begin{tabular}{c}
\begin{minipage}{6cm}
 \hspace{-1cm}
\includegraphics[scale=0.4]{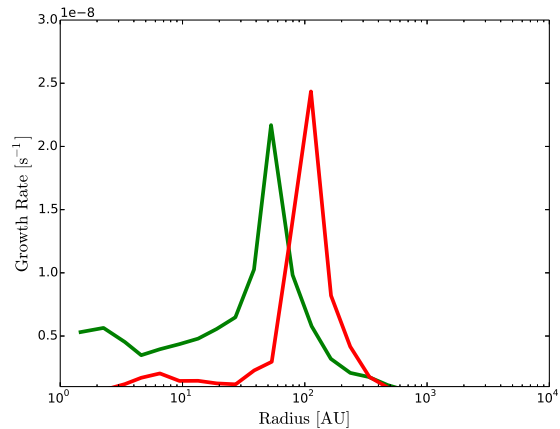}
\end{minipage} 
\end{tabular}
\caption{Spherically averaged positive growth rate of the magnetic field for a representative case (unsaturated case, halo A) is plotted against the radius for the earlier and later times. The green line shows the magnetic growth rate at the beginning of accretion shock while red line shows the growth rate close to the saturation stage. For the definition of growth rate see the text and references therein.}
\label{fign}
\end{figure*}

\begin{figure*}
\centering
\begin{tabular}{c}
\begin{minipage}{6cm}
 \hspace{-1cm}
\includegraphics[scale=0.4]{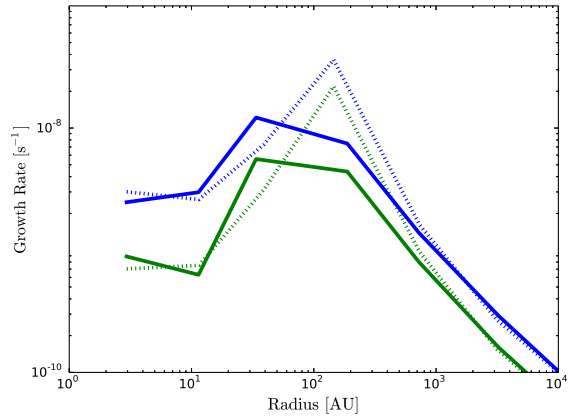}
\end{minipage} 
\end{tabular}
\caption{The absolute values of the growth rates of magnetic field amplification for a representative halo are plotted against the radius in this figure. The green color shows the magnetic growth rate due to the compression while blue color shows the  growth rate due to the shear. The dotted lines present unsaturated cases while the solid lines stand for the saturated cases. For the definition of growth rate see the text and references therein.}
\label{fign1}
\end{figure*}

\begin{figure*}
\hspace{-15.0cm}
\centering
\begin{tabular}{c}
\begin{minipage}{3cm}
\includegraphics[scale=1.5]{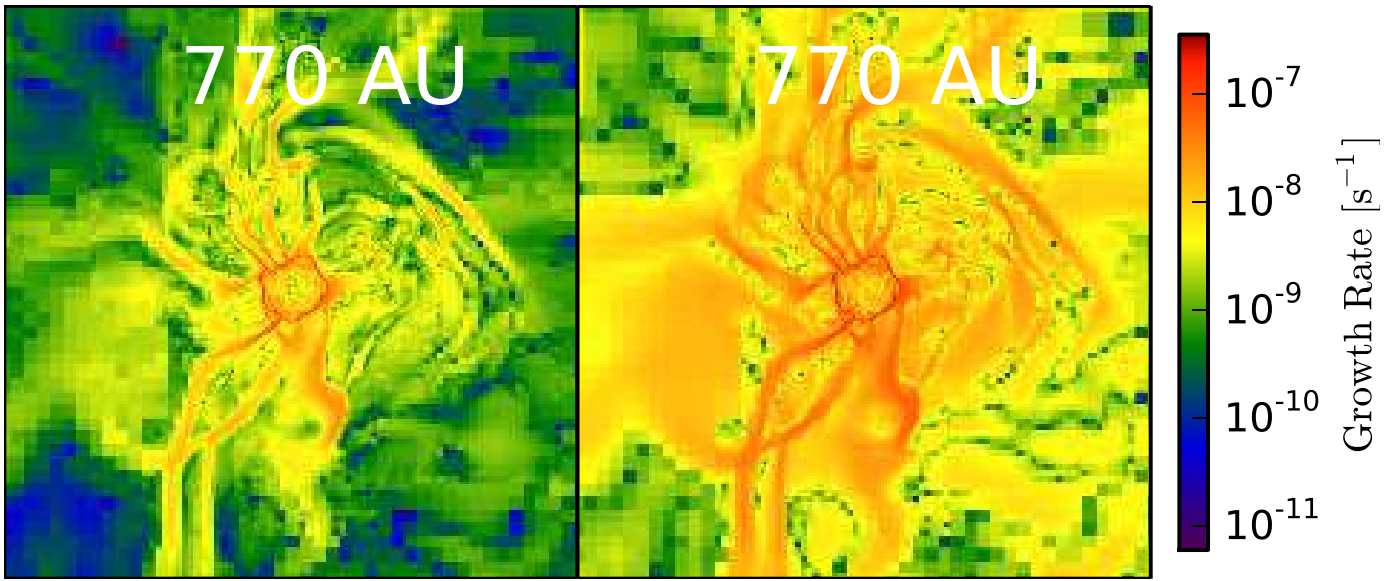}
\end{minipage}
\end{tabular}
\caption{This figure shows growth rates (absolute values) of magnetic field amplification by the shear (left) and compression (right) for unsaturated cases. The slices of growth rate are shown here centered at the peak density.}
\label{figh6}
\end{figure*}

\begin{figure*}
\centering
\begin{tabular}{c c}
\begin{minipage}{4cm}
\hspace{-4cm}
\includegraphics[scale=0.4]{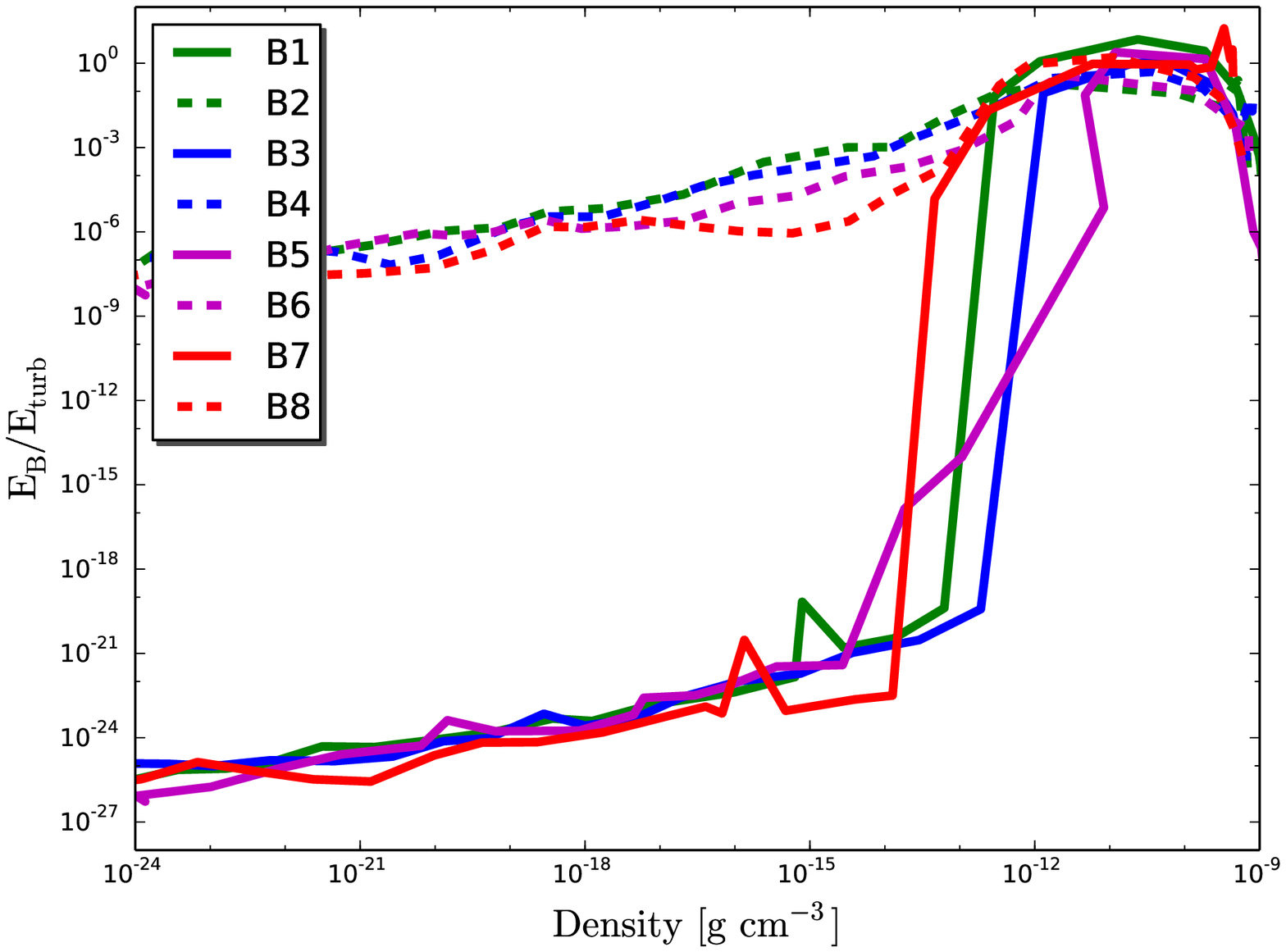}
\end{minipage} &
\begin{minipage}{4cm}
\includegraphics[scale=0.4]{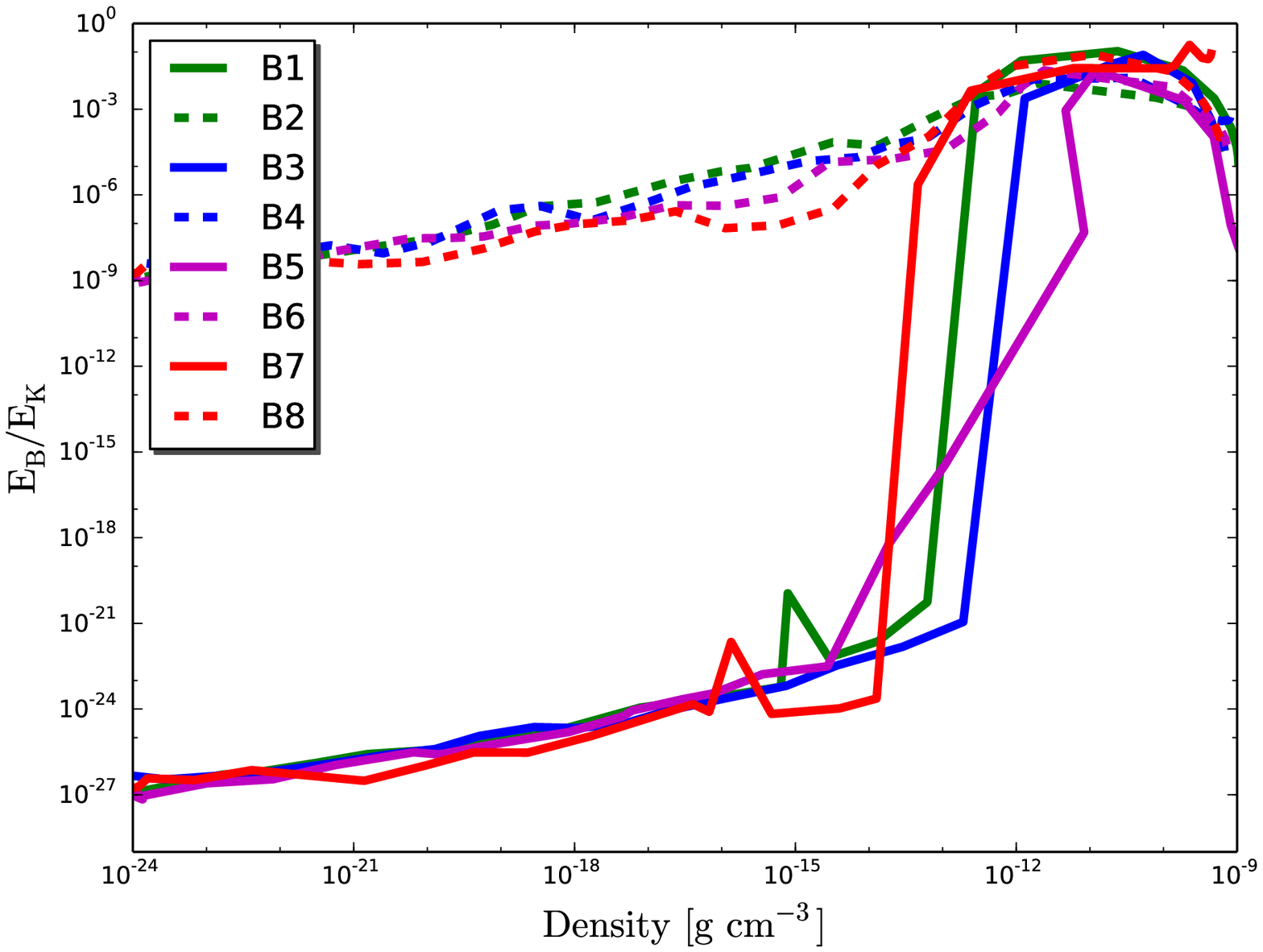}
\end{minipage} 
\end{tabular}
\caption{ The ratio of magnetic to turbulent energy (left panel) and magnetic to kinetic energy (right panel) is shown in the figure. B1, B3, B5, B7 represent the non-saturated cases while B2, B4, B6, B8 stand for the saturated field cases as listed in table \ref{table1}. }
\label{fig7}
\end{figure*}

\begin{figure*}
\centering
\begin{tabular}{c}
\begin{minipage}{6cm}
 \hspace{-1cm}
\includegraphics[scale=0.4]{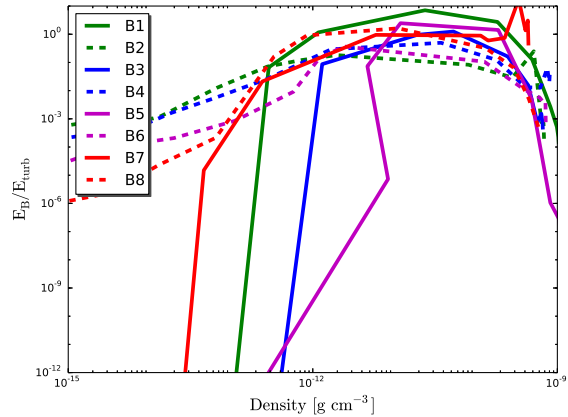}
\end{minipage} 
\end{tabular}
\caption{The ratio of magnetic to turbulent energy is shown in the figure for centeral region. Dashed and solid lines represent saturated and non-saturated cases as indicated in the legend.}
\label{fign2}
\end{figure*}

\begin{figure*}
\vspace{-1.0cm}
\hspace{-9.0cm}
\centering
\begin{tabular}{c}
\begin{minipage}{8cm}
\includegraphics[scale=0.8]{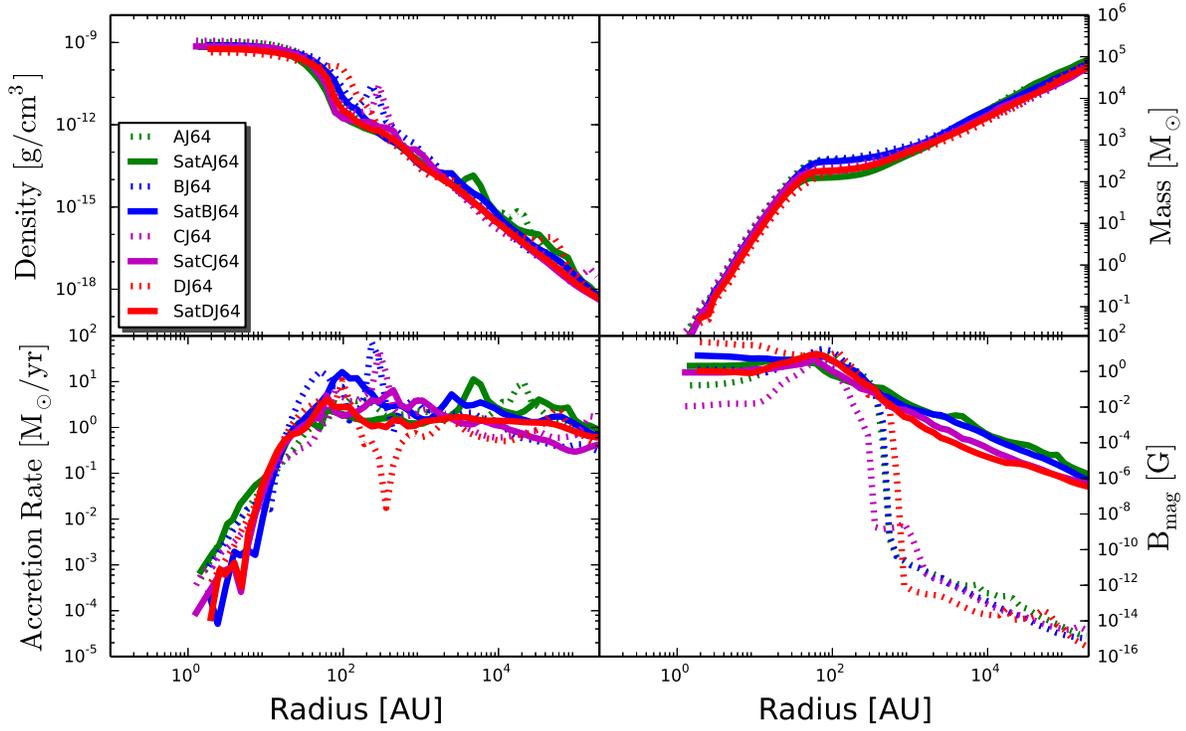}
\end{minipage}
\end{tabular}
\caption{ Radially binned spherically averaged radial profiles are shown for the halos A, B, C and D. The solid lines represent saturated cases while the dashed lines stand for non-saturated cases. Top left and right panels show the enclosed density and mass radial profiles. The accretion rates and magnetic field strength radial profiles  are depicted in the bottom left and right panels.}
\label{figh2}
\end{figure*}

\begin{figure*}
\hspace{-13.0cm}
\centering
\begin{tabular}{c}
\begin{minipage}{6cm}
\includegraphics[scale=0.22]{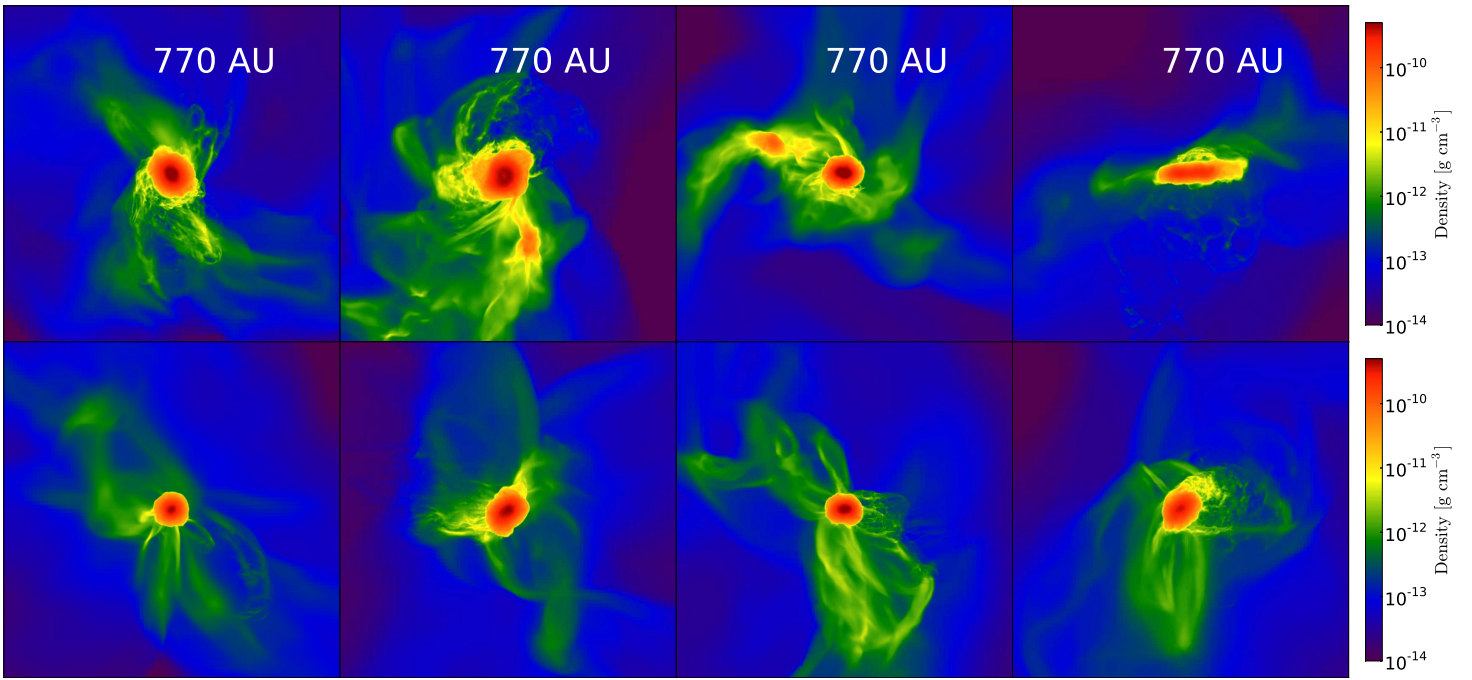}
\end{minipage}
\end{tabular}
\caption{The state of simulations is represented by the density-weighted mean along the axis of projection at the central peak density of $\rm 7 \times 10^{10}~g/cm^{3}$. Non-saturated cases (top panel) and saturated cases (bottom panel) are shown for halos A to D (starting from the left).}
\label{figh3}
\end{figure*}

\begin{figure*}
\hspace{-13.0cm}
\centering
\begin{tabular}{c}
\begin{minipage}{6cm}
\includegraphics[scale=0.22]{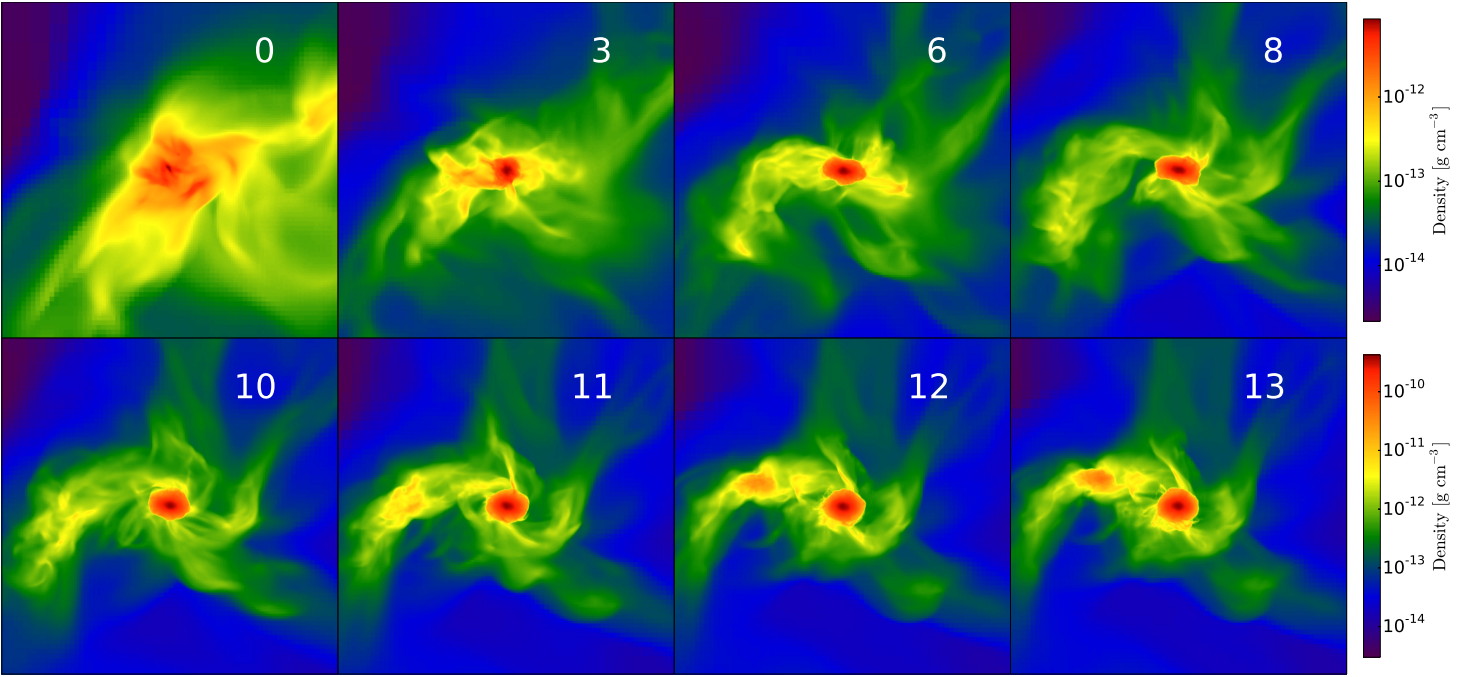}
\end{minipage}
\end{tabular}
\caption{The time evolution of the density-weighted mean along the axis of projection for the halo C. The time in years, after the formation of the first peak is shown in each case for the central 770 AU.}
\label{figh31}
\end{figure*}

\begin{figure*}
\centering
\begin{tabular}{c c}
\begin{minipage}{4cm}
\hspace{-4cm}
\includegraphics[scale=0.4]{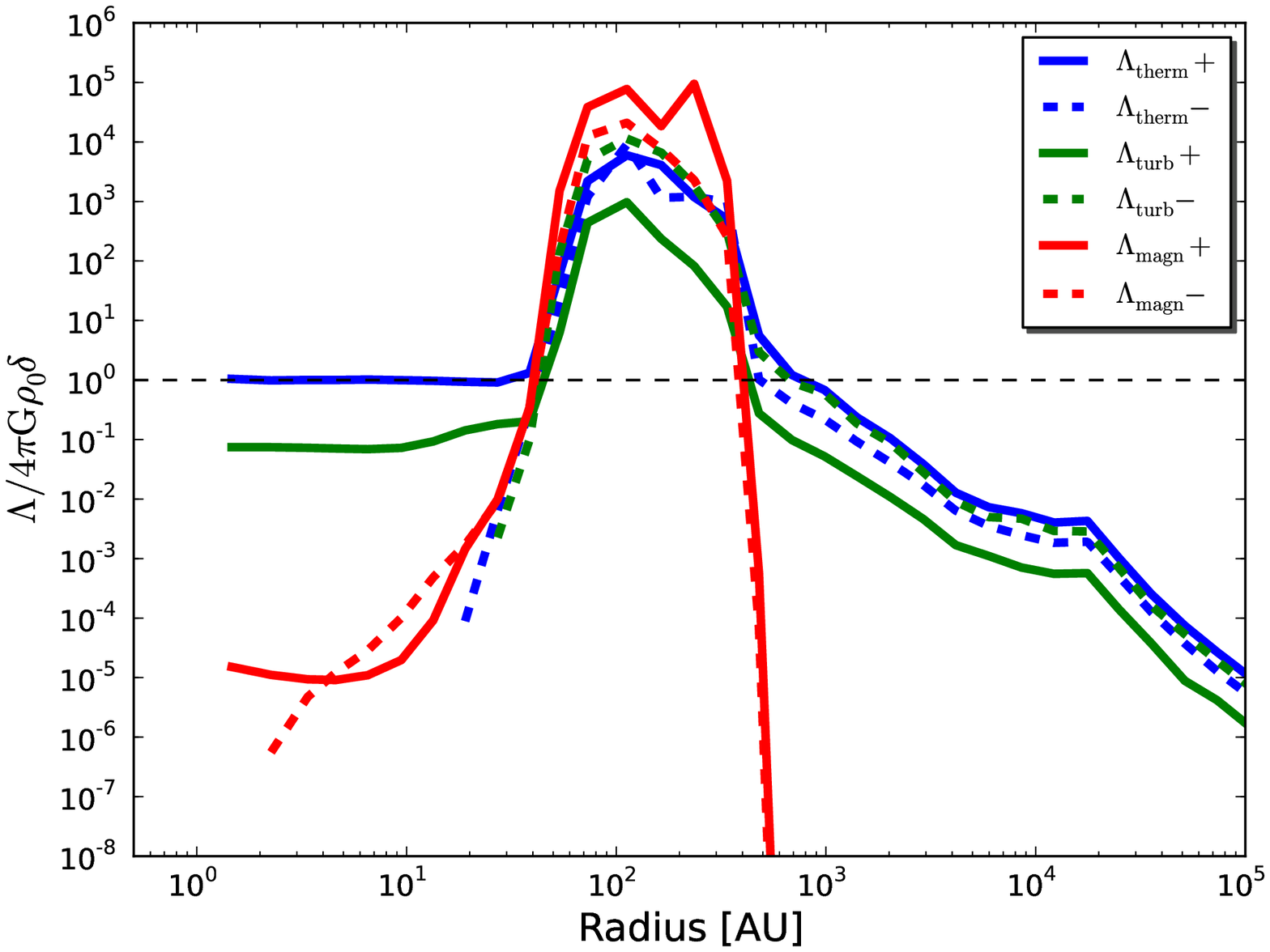}
\end{minipage} &
\begin{minipage}{4cm}
\includegraphics[scale=0.4]{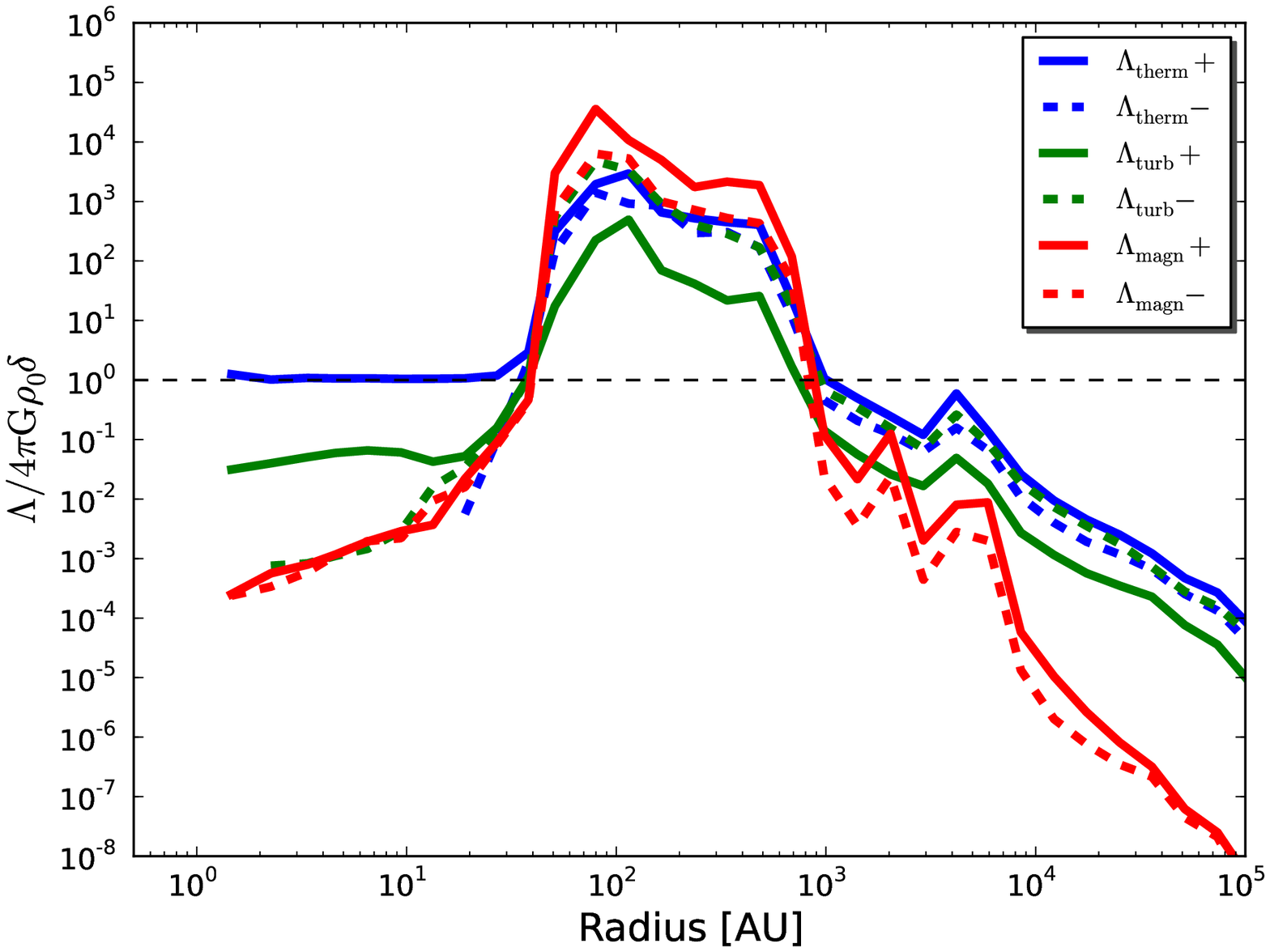}
\end{minipage} \\
\begin{minipage}{4cm}
\hspace{-4cm}
\includegraphics[scale=0.4]{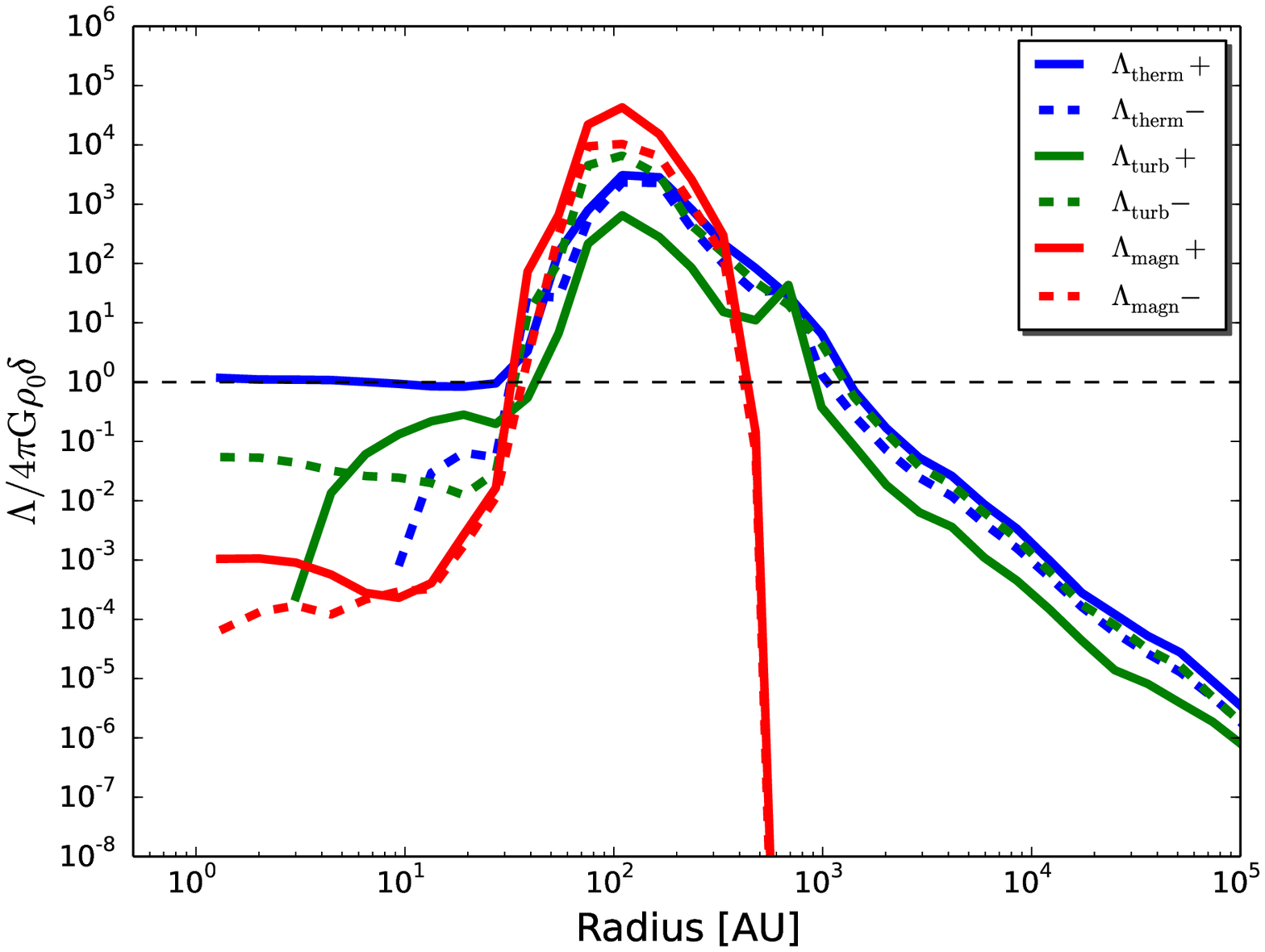}
\end{minipage} &
\begin{minipage}{4cm}
\includegraphics[scale=0.4]{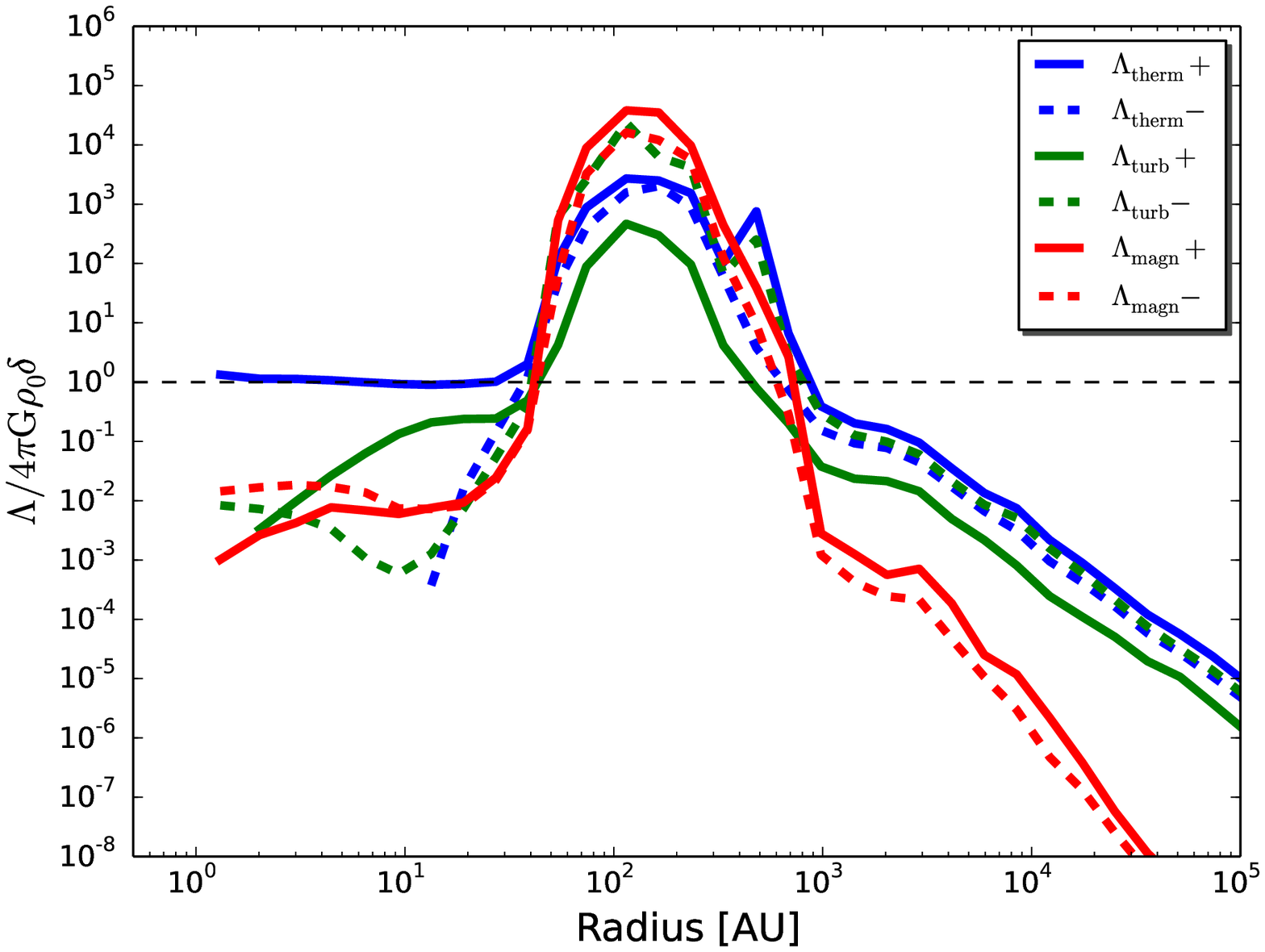}
\end{minipage} 

\end{tabular}
\caption{Figure shows the contribution of support terms for both saturated (right) and non-saturated (left) cases. The upper and lower panels represent two different halos. The solid lines represent the positive support of the quantities while dashed lines represent the negative support. For the definitions of support terms see text. The local support of thermal, turbulent and magnetic fields is shown in this figure.}
\label{fig9}
\end{figure*}

\section{Main Results}

In all, we have performed 8 cosmological magnetohydrodynamics simulations for four distinct halos each with an initial magnetic field strength of $\rm 3 \times 10^{-20}$ G (hereafter called non-saturated cases) and $\rm 3 \times 10^{-11}$ G (hereafter called saturated cases). As shown by \cite{2013MNRAS.432..668L},  the latter implies an approximate equipartition between magnetic and turbulent energy at densities of $\rm 10^{-12}~g/cm^{3}$. The lower value is characteristic for magnetic field generation via the Biermann battery \citep{1950ZNatA...5...65B} or through thermal fluctuations \citep{2012PhRvL.109z1101S}, while the higher value may occur if magnetic fields are generated during the QCD or electroweak phase transition \citep{2004PhRvD..70l3003B}. The initial masses and collapse redshifts of the halos are listed in table \ref{table1}. The density perturbations decouple from the Hubble flow and start to collapse via gravitational instability. The gas falls into the dark mater potential and gets shock heated. This process continues until the gas temperature exceeds $10^4$~K where cooling due to Lyman alpha radiation becomes effective and brings the temperature down to 8000 K. Further cooling to lower temperatures remains suppressed due to the photo-dissociation of $\rm H_{2}$ molecules by the strong Lyman-Werner flux. Consequently, an isothermal collapse occurs. The density profile follows an $R \rm ^{-2}$ behavior as expected from an isothermal collapse. The radial infall velocity is about 10 $\rm km/s$. Overall, the collapse dynamics is similar to our previous studies during the initial phases \citep{2013MNRAS.432..668L,2013MNRAS.433.1607L}. In the following, we explore the amplification of magnetic fields during the collapse and their impact on fragmentation.

\subsection{Amplification of magnetic fields}

The simulations were started with initial seed magnetic fields as mentioned in table \ref{table1}. For both runs, the magnetic field is mainly amplified by gravitational compression below densities of $\rm 10^{-12}~g/cm^{3}$. In this regime, the amplification of the magnetic field strength scales with $B \propto \rho^{2/3}$ and the strength of the magnetic field  at the end of our simulations remains much weaker in the non-saturated runs compared to the saturated cases as shown in figure \ref{figh1}. It is found that the strength of the magnetic fields becomes almost equal at densities of  $\rm 7 \times 10^{-10}~g/cm^3$ for both the saturated and non-saturated cases during the transition to the adiabatic evolution after reaching the maximum refinement level. This is evident from figure \ref{figh1}. As we will show in the following, this rapid amplification in the non-saturated runs is due to the occurrence of strong accretion shocks. At densities above $\rm 10^{-11}~g/cm^3$, the evolution becomes adiabatic, and a stable core  is formed which reaches the state of hydrostatic equilibrium. This core is considered as a proxy for a supermassive protostar. The infall of the gas onto the central core results in the formation of accretion shocks.

To investigate the rapid amplification of non-saturated magnetic fields in accretion shocks, we show the time evolution of the magnetic field strength and the radial velocity profiles for the two representative cases in figures \ref{fig2} and \ref{fig3}. It can be noted from the figures that the amplification of magnetic fields is closely related to the radial infall velocity. The radial velocity increases from 10 $\rm km/s$ to 30 $\rm km/s$ (even higher) within a time scale of about 1 year.  Similarly, the field strength is amplified by a few orders of magnitude during the same time. The profile of the radial infall velocity is smooth in the beginning and becomes sharper as accretion shocks are formed. The sharp jump in the radial velocity profile around 100 AU is a typical signature of the accretion shocks. It may further be noted that the increase in density corresponding to different times in figures \ref{fig2} and \ref{fig3} is just about an order of magnitude. Due to gravitational compression B should increase with $\rho^{2/3}$, which could not explain such large increase in magnetic field strength. Amplification by gravitational compression would further be more homogenous with in the Jeans volume. Thus, amplification is due to the accretion shocks.

As demonstrated by figure \ref{fig2}, the initial amplification occurs at the shock front enclosing the core and subsequently grows inside the core until the end of the simulation is reached. Apart from turbulent diffusion, amplification by shear can contribute to the growth of the magnetic field in the core. Additional contributions may also come from the advection of the magnetic field in the core and diamagnetic pumping from a gradient in the turbulent intensity.

 
In order to understand the different contributions to the magnetic field amplification, we have computed the growth rates of amplification both by turbulent shear and compression for all halos. The rate of change of the magnetic pressure in a fluid element moving with the flow can be computed from the  source terms of the induction equation \citep{2013MNRAS.431.3196S}:
\begin{equation}
 {D \over Dt} \left( {B^{2} \over 8 \pi} \right) =  {1 \over 4 \pi}\left(B_{i}B_{j}S_{ij}^{*} - {2 \over 3}B^{2}d \right),
\label{Bpres}
\end{equation}
where $\rm{D \over Dt}$ is $\rm {\partial \over \partial t} + v \cdot \nabla$, $d=\nabla \cdot v$ is the velocity divergence and $S_{ij}^{*}=S_{ij}-1/3 d\delta_{ij}$ is the trace-free rate of strain tensor. Dividing both sides of equation (\ref{Bpres}) by $B^2/8\pi $, the first term  on the right-hand side represents the growth rate of magnetic energy by shear while the second term is the growth rate by compression (both due to gravity and shocks). The absolute values of the growth rates by shear and compression are shown in figure \ref{fig5}. They increase towards smaller radii, peak around 100 AU and decline toward the center. Such a trend is observed for all halos, both for the saturated and the non-saturated cases. To further elucidate the differences in the amplification rates for the saturated and unsaturated cases, we have overplotted the amplification rates by the shear and compression for a representative case in figure \ref{fign1}. The plot shows that the growth rate is higher for the unsaturated case. It is also noted that the growth rate is higher for shear compared to the compression. In figure \ref{fign}, we show the spherically averaged positive growth rate for a representative case (i.e., halo A, unsaturated run) at the start of the accretion shock and close to the saturation state. The strongest amplification occurs at the accretion shock on scales of a few 100 AU. When saturation occurs in the central region, the growth rate declines in the core and the peak in the growth rate shifts towards larger radii. This is a clear indication of magnetic field saturation on small scales while amplification still proceeds on scales larger than 100 AU. The inverse of the growth rate gives the amplification time scale which locally decreases to less than 0.1 year. The amplification time scale for compression is about half of the shear amplification time scale. We note here that even for compressively driven turbulence part of the energy (about $\rm 1/3-1/2$) lies in the solenoidal modes \citep{2010A&A...512A..81F} thus naturally providing a two-to-one ratio of compressive and shear modes. The local variations in the growth rates are shown in figure \ref{figh6} for a representative case. The  very short amplification time scale shows that the magnetic field can be amplified very rapidly in the presence of strong accretion shocks. In the center of the core, however, the amplification by compression and shear is weak. This suggests that the growth of the magnetic field inside the core (see figures 2 and 3) is mainly caused by advection and turbulent diffusion.

To further assess the amplification and saturation of magnetic fields, we have computed the ratio of the magnetic to turbulent energy as shown in figure \ref{fig7}. It increases with density both for the saturated and the non-saturated cases. At the strong accretion shock the magnetic field amplification time scale becomes very short and rapid amplification happens for both saturated and non-saturated cases until the magnetic energy becomes comparable the turbulent energy. To further clarify the differences between the saturated and unsaturated cases, we have plotted the ratio of magnetic to turbulent energy for the centeral region. This figure shows that magnetic energy is in equipartition with turbulent energy and the magnetic field gets saturated as evident from the change in the slop of $\rm E_{B}/E_{turb}$. The ratio of the magnetic to the total kinetic energy is depicted in the right panel of figure \ref{fig7}. It initially increases with density, gets enhanced rapidly by accretion shocks, reaches a peak value of $\rm 10^{-1}$ and then declines which is an indication of magnetic field saturation. It is further noted that saturation occurs around densities of $\rm 10^{-10}~g/cm^{3}$ which is deduced from the change in the slope of the magnetic to kinetic energy ratio. In the saturated cases, the amplification is  many orders of magnitude lower compared to their counterparts which reach the same field strength from a much smaller value. This is expected, as the initial seed field is already in approximate equipartition with the turbulent energy, implying amplification predominantly by gravitational compression.  


\subsection{Implications for the formation of seed black holes}

The central properties of the halos at their collapse redshifts are shown in figure \ref{figh2}. The density profile shows an $R^{-2}$ behavior as expected from an isothermal collapse and becomes flat in the central adiabatic core. This trend is observed for all cases. The small bumps in the density profiles for non-saturated cases are due to the formation of additional clumps. The maximum  value of the density is $\rm 7 \times 10^{-8}~g/cm^{3}$. The mass profile increases with $R^{2}$ in the center, becomes flat around 100 AU and then increases linearly with radius. The mass profiles are very similar for the saturated and non-saturated cases. The mass accretion rates are about 1 $\rm M_{\odot}/yr$ at larger radii and drop down to  $\rm 10^{-3}~M_{\odot}/yr$ in the central adiabatic core. The profile of the magnetic field strength shows that irrespective of the initial seed field, the magnetic field reaches the saturation value in the presence of strong accretion shocks. Overall, the halo properties are in good agreement with previous studies \citep{2013MNRAS.433.1607L,2013arXiv1309.1097L}.

The state of the simulations is shown by the density-weighted mean along the projection axis for four distinct halos  in figure \ref{figh3}. It is found that massive clumps of a few hundred solar masses are formed in every halo both for the saturated and the non-saturated runs. In addition to this, fragmentation is observed in two halos for the non-saturated cases. The masses of these clumps are a few ten solar masses (20 $\rm M_{\odot}$ and 30 $\rm M_{\odot}$) and they are gravitationally bound. The suppression of fragmentation in the saturated cases is attributed to the additional magnetic pressure on larger scales. The time evolution of the density structure for halo C is shown in figure \ref{figh31}. The initially turbulent cloud collapses to form a massive clump within a few years. It keeps accreting gas from its surroundings and the formation of an additional clump can be seen after 10 years of evolution. 

We have investigated the impact of magnetic fields on the fragmentation properties of these halos and computed the local support terms for magnetic fields. The local support is derived from the source terms of the differential equation for the rate of compressions of the gas \citep{2013MNRAS.431.3196S}
\begin{equation}
-{D d \over Dt}= 4 \pi G \rho_{0}\delta -\Lambda
\end{equation}
Here, $\delta$ is the overdensity relative to the mean density $\rho_0$ and 
$\Lambda$ is the local support against gravitational compression. $\Lambda$ receives the contributions from thermal pressure, resolved turbulence, and the magnetic fields. The support by magnetic fields is \citep{2013MNRAS.431.3196S}:
\begin{dmath}
\Lambda_{\rm mag} = {1 \over 4 \pi \rho} \left[-{ \partial^{2} \over \partial x_{i} \partial x_{j}} \left( {1 \over 2} B^{2} \right) + {\partial B_{i} \over \partial x_{j}} {\partial B_{j} \over \partial x_{i}} \right] + \\
 {1 \over 4 \pi \rho^{2}} {\partial \rho \over \partial x_{i}} \left[ { \partial \over \partial x_{i}} \left( {1 \over 2} B^{2} \right) + B_{j}{\partial B_{i} \over \partial x_{j}} \right] 
\end{dmath}
For the definition of thermal and turbulence support terms see \cite{2013MNRAS.431.3196S}, while a first application is presented by \cite{2013MNRAS.433.1607L}. Like other support terms, the magnetic support has positive and negative components. The positive components provide support against gravity while the negative components aid to the gravitational compression. The contributions of the local support terms against gravity are shown for two representative cases in figure \ref{fig9}. It is important to note that the contribution of the positive support by the magnetic field dominates over the turbulent and thermal pressure support in the vicinity of the accretion shocks at radii around 100 AU. The support by turbulence is dominated by the negative contribution from compression by accretion shocks. Negative support is a characteristic of compressible turbulence, particularly in the presence of shocks. For a detailed discussion on negative turbulent support see \cite{2013MNRAS.431.3196S}. Even stronger support comes from the thermal pressure, while the magnetic support is sub-dominant near the center. For the saturated field cases, the large positive support from magnetic fields helps in the suppression of fragmentation on radial scales ranging from less than 100 to about 1000 AU, which encompasses the fragmentation scale in figure \ref{figh3}. Particularly at radii outside the accretion shock, this is a result of the initially larger magnetic field. For the unsaturated case, the amplification of the magnetic field produces support comparable to the saturated case only for a narrower range of scales around 100 AU. As numerical simulations tend to underestimate the physical amplification rate, a final conclusion on the role of initial field strength and dynamo support is, however, not possible at this stage.


\section{Discussion}

In total, we have performed 8 cosmological MHD simulations to investigate the role of magnetic fields during the formation of supermassive black holes. The simulations were carried out for four distinct halos with initial seed magnetic fields of $\rm 3 \times 10^{-20}$ G  and $\rm 3 \times 10^{-11}$ G. The main motivation for the selection of the stronger magnetic field strength was to explore the impact of saturated magnetic fields on the fragmentation properties of so-called atomic cooling halos. To achieve this goal, we evolved the simulations adiabatically beyond the formation of the first peak for a few free-fall times until they reached the same peak density of $\rm 7 \times 10^{-10}~g/cm^{3}$. 

Our results show that irrespective of the initial seed field strength, the magnetic field gets amplified very rapidly in the presence of strong accretion shocks. This is indicated by the short time scale for compressive amplification compared to the free-fall time. The amplification is mainly caused by the shock fronts and the magnetic field is subsequently transported into the core by turbulent diffusion and similar processes until the magnetic energy grows to equipartition with kinetic energy. We therefore report a new mode of  magnetic field amplification by the accretion shocks in atomic cooling halos as well as a possible contribution from the compressive turbulent modes driven by the accretion process. 

We further note that, while the adiabatic cores in our simulations were introduced to follow the collapse beyond the first peak, very similar cores are expected to form during the formation of protostars at higher densities \citep{2001ApJ...546..635O,2008ApJ...686..801O}. It is thus desirable to extend the calculations pursued here to the formation of protostars. We also emphasize that the turbulent amplification of magnetic fields depends strongly on the Reynolds number of the flow \citep{1968JETP...26.1031K,1998MNRAS.294..718S}. Since we cannot resolve all length scales down to the physical dissipation length scale, the actual amplification is probably even stronger. Such rapidly amplified magnetic fields may suppress the fragmentation on even larger scales than shown in these simulations. Currently, however, fully resolved simulations are infeasible. A possible solution to this problem might be the application of subgrid-scale models for MHD turbulence. 

Our results indicate that magnetic fields are relevant for the formation of seed black holes, as they help in the suppression of fragmentation via additional magnetic pressure. The masses of the clumps at the end of our simulations are a few hundred solar masses and large accretion rates of about $\rm 1~M_{\odot}/yr$ are observed. Given such high accretion rates, these objects are expected to reach $\rm 10^5~M_{\odot}/yr$ within a short time. The amount of fragmentation is significantly less compared to the hydrodynamics simulations performed in our previous study  \citep{2013MNRAS.433.1607L}. The peak density reached in the MHD simulations is about 13 times lower than in the hydrodynamical case. Further differences may arise from the use of different Riemann solvers.

We evolved these simulations only for a few free fall times after the formation of the first peak. Further evolution of such high-resolution simulations becomes extremely costly due to the Courant constraints on the computation of timestep. However, we expect that the presence of the magnetic fields will be favorable for the formation of massive seed black holes as it suppresses the fragmentation. We have also shown in recent studies that subgrid scale turbulence helps in the formation of stable accretion disks and assembling massive objects of $\rm 10^5~M_{\odot}$ in 20,000 years via rapid accretion \citep{2013MNRAS.433.1607L,2013arXiv1309.1097L}. The presence of subgrid scale MHD turbulence may further help in the formation of accretion disks in magnetized halos. As our previous results indicated that accretion stalls when $\rm  \sim 10^{5}~M_{\odot}$ are reached because of an increase in the rotational support, we speculate that magnetic fields may enhance angular momentum transport and increase the final mass scale. This requires cosmological MHD simulations following the accretion for even longer times.

\section*{Acknowledgments}
The simulations described in this work were performed using the Enzo code, developed by the Laboratory for Computational Astrophysics at the University of California in San Diego (http://lca.ucsd.edu). We acknowledge research funding by Deutsche Forschungsgemeinschaft (DFG) under grant SFB $\rm 963/1$ (projects A12, A15) and computing time from HLRN under project nip00029. DRGS thanks the DFG for funding via the Schwerpunktprogram SPP 1573 ``Physics of the Interstellar Medium'' (grant SCHL $\rm 1964/1-1$). The simulation results are analyzed using the visualization toolkit for astrophysical data YT \citep{2011ApJS..192....9T}.

\bibliography{blackholes.bib}
\end{document}